\documentclass[10pt,aps,prd,showpacs,footinbib,twocolumn]{revtex4-1}
\usepackage[usenames,dvipsnames]{color}
\usepackage{amsfonts}
\usepackage{amssymb}
\usepackage{amsmath}
\usepackage{graphicx}
\usepackage{epstopdf}
\usepackage[colorlinks]{hyperref}
\usepackage[normalem]{ulem}

\begin{document}

\title{Highly compact neutron stars and screening mechanisms: \\  Equilibrium and stability}

\author{Bernardo F.\ de Aguiar}
\email{bernardo\_aguiar@id.uff.br }
\affiliation{Instituto de F\'isica, Universidade Federal Fluminense, Niter\'oi, 
Rio de Janeiro, 24210-346, Brazil.}
\author{Raissa F.\ P.\ Mendes}
\email{rfpmendes@id.uff.br}
\affiliation{Instituto de F\'isica, Universidade Federal Fluminense, Niter\'oi, 
Rio de Janeiro, 24210-346, Brazil.}

\date{\today}

\begin{abstract}
Modified theories of gravity that offer viable models for dark energy often rely on mechanisms that screen their effects in high density environments. From this perspective, it would appear that, once solar system constraints are satisfied, these theories would predict a trivial phenomenology for (much denser) neutron stars. In this work we explore the fact that in scalar-tensor theories the scalar degree of freedom does not couple to the mass density alone, but to the trace of the energy-momentum tensor---which can increase and eventually change sign as density and pressure build up in the core of neutron stars---, and investigate whether there could be a partial unscreening of the scalar field inside the most compact stars found in Nature. 
For this purpose, we construct neutron star solutions with realistic equations of state in theories with screening mechanisms and study their stability under radial perturbations. In particular, we find that stable solutions with an unscreened core can exist in chameleon models, while for the environmentally-dependent dilaton model a wealth of new, scalarized equilibrium solutions are found, some of which can be stable.
\end{abstract}


\maketitle

\section{Introduction}

In order to offer viable models for the late-time accelerated expansion of the Universe, modified theories of gravity must pass solar system and other astrophysical tests, which are in excellent agreement with general relativity (GR) \cite{Clifton2012,Joyce2014,Berti2015}. This is often accomplished through the so-called screening mechanisms, which suppress possible fifth forces mediated by the new degrees of freedom in high density environments, while unleashing their effects at cosmological scales.
Incarnations of the screening mechanism include scalar-tensor theories of the chameleon type \cite{Khoury2004a,Khoury2004}, which rely on the environmental dependence of the scalar field effective mass, the symmetron model \cite{Hinterbichler2010}, where decoupling is achieved through a symmetry breaking potential, the environmentally-dependent dilaton model \cite{Brax2010}, as well as kinetic-type screening, such as the Vainshtein mechanism \cite{Vainshtein1972,Babichev2013}.

A broad class of modified theories of gravity displaying screening effects contains an additional scalar degree of freedom which obeys an equation of the type 
$\Box \phi = V_{\textrm{eff},\phi}(\phi)$,
where $V_\textrm{eff}$ depends both on the scalar field potential $V(\phi)$ and on its coupling to matter $A(\phi)$. In the Newtonian limit, $V_\textrm{eff} (\phi) = V(\phi) + \rho \ln A(\phi)$, where $\rho$ is the matter rest-mass density. The essence of a screening mechanism is then to choose the model functions $V(\phi)$ and $A(\phi)$ in such a way that the scalar field is hidden or suppressed in high density regions, where astrophysical constraints are tighter, while remaining unfettered in the low density environments relevant to cosmology.

Once a model presenting a screening mechanism is tuned to general relativity for typical solar system densities, it would seem to follow that no new phenomenology would arise in much denser environments. 
However, in a relativistic scenario the scalar field equation actually implies a coupling to the trace $T$ of the energy-momentum tensor of matter fields, and not to the mass density alone: $V_\textrm{eff} (\phi) = V(\phi) - T \ln A(\phi)$. For a perfect fluid, $T = 3p - \epsilon$, where $p$ is the pressure and $\epsilon$ is the energy density in the fluid's rest frame. Although $T$ is typically dominated by the rest-mass contribution to the energy density, $T \approx - \rho$, this fails to hold in the core of the densest objects in Nature: neutron stars (NSs). 

Although the equation of state (EoS) describing the microscopic behavior of matter inside NSs is still not fully understood \cite{Haensel2007,Ozel2016}, measurements of NS properties suggest that matter at several times the nuclear saturation density may display intriguing properties. The high masses of some observed NSs indicate that in their core the EoS should be relatively stiff, with a large speed of sound \cite{Bedaque2015,Alsing2018}. An interesting possibility, that concerns us here, is the conceivable appearance of a pressure-dominated phase in the core of the most compact NSs in Nature, such that $p>\epsilon/3$ (and $T>0$) in a region of their interior \cite{Mendes2015,Podkowka2018}.
If that is the case, the screening mechanisms thought to effectively suppress fifth force effects in high density environments could (at least partially) fail where densities are the highest. In this work we explore this possibility. 

In the first part of this paper, we study the structure of neutron stars in models of screened modified gravity, using realistic EoS that allow for a pressure-dominated phase in their core. We thus revisit the works of Refs.~\cite{Babichev2010,Brax2017} with a more realistic description of the NS interior and a more thorough analysis of the space of solutions. We find that the scalar field profile can differ radically between NS configurations with and without a pressure-dominated core, and that this can leave imprints on global quantities such as the neutron star mass. The two models discussed in this work, namely the chameleon and the environmentally-dependent dilaton models, illustrate the diverse phenomenology displayed by NSs with pressure-dominated cores.
While in the chameleon model the main effect is a change of the scalar field profile and suppression of the thin-shell effect, in the environmentally-dependent dilaton model we find a wealth of new equilibrium solutions, which were missing in previous analyses \cite{Brax2017}. 
Indeed, the space of solutions in the environmentally-dependent dilaton model resembles what was found in the context of the spontaneous scalarization of a massless nonminimally coupled scalar field with a positive nonminimal coupling \cite{Mendes2016}, and widens the range of interesting theories displaying this scalarization effect.

As our second main contribution, we consider the general problem of linear radial perturbations around the equilibrium configurations found initially. 
This serves a three-fold purpose. First, in Refs.~\cite{Babichev2010,Brax2017}, questions were raised about the stability of equilibrium solutions with pressure-dominated interiors. The stability issue is clarified through our analysis. We find evidence that stable, partially unscreened solutions exist, and that dynamical instability sets in at the turning point of a sequence of equilibrium solutions, as in General Relativity.
Second, the stability analysis helps to clarify the physical significance of the many scalarized solutions found in the environmentally dependent dilaton model. 
Third, the study of radial perturbations can be seen as a first step towards the investigation of generic NS oscillation modes in models of screened modified gravity. Oscillation modes carry information about the NS interior \cite{Baiotti2019}, and provide a promising observational tool to access the scalar field activation that might occur in the core of highly compact neutron stars. This last avenue will be further pursued in a future work.

This paper is organized as follows. In Sec.~\ref{sec:framework}, we present the general framework under consideration, including the gravity models and NS EoS that will be employed. In Sec.~\ref{sec:equations} we display  the structure equations describing static and spherically symmetric NSs in models of screened modified gravity, and derive the equations governing linear radial perturbations around these solutions. In Sec.~\ref{sec:results} we present a collection of results for NSs in the chameleon and the environmentally dependent dilaton models, and investigate their equilibrium and stability properties. Sec.~\ref{sec:conclusion} is devoted to final comments and conclusions. In what follows, we adopt natural units in which $c = G = \hbar = 1$ unless stated otherwise; also $M_\textrm{Pl} = \hbar c / \sqrt{8 \pi G}$ denotes the (reduced) Planck mass.

\section{Framework} \label{sec:framework}

\subsection{Field equations}

Various models of modified gravity exhibiting screening mechanisms---such as chameleons, symmetrons, dilatons, and $f(R)$---can be described in a unified framework through the action \cite{Brax2012}
\begin{align} \label{eq:generalaction}
S & =\int d^4 x \sqrt{-g}
\left[ \frac{R}{16\pi}  - \frac{1}{2} g^{\mu \nu} \nabla_{\mu} \phi \nabla_{\nu} \phi - V(\phi) \right]
\nonumber \\
& + S_{m}[\Psi_m ; A(\phi)^2 g_{\mu\nu}],
\end{align}
where $\Psi_m$ denotes the collection of matter fields. These models include one additional scalar field, with a potential $V(\phi)$ and which couples to matter through the conformally rescaled (Jordan-frame) metric $\tilde{g}_{\mu\nu} \equiv A(\phi)^2 g_{\mu\nu}$. By choosing the two free functions $V(\phi)$ and $A(\phi)$ one fixes a particular model in this class.

The field equations obtained through the variation of Eq.~(\ref{eq:generalaction}) with respect to the metric and scalar field are given by
\begin{align} 
&G_{\mu \nu} = 8\pi \left[ T_{\mu \nu } + \nabla_{\mu} \phi\nabla_{\nu} \phi - g_{\mu \nu} \left (\frac{1}{2}\nabla^{\beta}\phi\nabla_{\beta}\phi + V(\phi) \right ) \right], \label{eq:field1}\\ 
&\nabla^\mu \nabla_\mu \phi = \frac{d V_\textrm{eff} }{d \phi},\label{eq:field2}
\end{align}
where the effective potential $V_\textrm{eff} (\phi)$ is defined as
\begin{equation}\label{eq:Veff}
V_\textrm{eff}(\phi) \equiv V(\phi) - T \ln A(\phi),
\end{equation}
with $T \equiv g^{\mu\nu} T_{\mu\nu}$ and the energy-momentum tensor of matter fields given by 
\begin{equation}
T_{\mu\nu} \equiv -\frac{2}{\sqrt{-g}} \frac{\delta S_m}{\delta g^{\mu\nu}}. 
\end{equation}

By taking the divergence of Eq.~(\ref{eq:field1}), one obtains the matter equations of motion,
\begin{equation}\label{eq:eqmotion}
\nabla^\nu T_{\mu\nu} = \frac{d \ln A}{d\phi} T \nabla_\mu \phi.
\end{equation}

Equation (\ref{eq:eqmotion}) reveals that test particles describe trajectories which are not geodesics of the (Einstein-frame) metric $g_{\mu\nu}$, but are instead forced by the scalar field gradient. Indeed, if $u^\mu$ is the particle's four-velocity, the scalar-induced acceleration is given by
\begin{equation} \label{eq:accel}
a^\mu \equiv u^\nu \nabla_\nu u^\mu = - P^{\mu\nu} \partial_\nu \ln A,
\end{equation}
where $P^{\mu\nu}\equiv g^{\mu\nu} + u^\mu u^\nu$ is the projector operator onto the subspace orthogonal to $u^\mu$.

In an alternative description in terms of the conformally related metric $\tilde{g}_{\mu\nu} \equiv A(\phi)^2 g_{\mu\nu}$, one could define the Jordan-frame energy-momentum tensor as $\tilde{T}_{\mu\nu} \equiv -2 (-\tilde{g})^{-1/2} \delta S_m/ \delta \tilde{g}^{\mu\nu} = A(\phi)^{-2} T_{\mu\nu}$. From Eq.~(\ref{eq:eqmotion}) it is easy to show that the Jordan-frame energy-momentum tensor is covariantly conserved: $\tilde{\nabla}^\nu \tilde{T}_{\mu\nu} = 0$, and free particles follow geodesics of $\tilde{g}_{\mu\nu}$.

\subsection{Screening mechanism}

Screening mechanisms aim to suppress the acceleration (or fifth force) given by Eq.~(\ref{eq:accel}) in solar system (or galaxy) scales, where GR is very well tested. To understand on general grounds how screening works, let us assume that the scalar field has settled, on a larger scale, at a constant value $\phi = \phi_0$, which, by Eq.~(\ref{eq:field2}), should correspond to a minimum of the effective potential: $dV_\textrm{eff}/d\phi |_{\phi_0}= 0$. In this background, let us consider a spherical body with mass density distribution $\rho$, total mass $M$, and radius $R$. Since the scalar field is sourced by matter, the field profile will change around the massive object, giving rise to a nonzero fifth force. 
Setting $\phi = \phi_0 + \delta \phi$, and assuming that gravity is weak ($M/R \ll 1$), we can approximate the scalar field equation (\ref{eq:field2}) to linear order by
\begin{equation}\label{eq:Newtonian}
\nabla^2 \delta \phi - m_\textrm{eff}^2 (\phi_0) \delta \phi = \rho \left. \frac{d \ln A}{d\phi} \right|_{\phi_0} ,
\end{equation}
where $m_\textrm{eff}^2 \equiv d^2 V_\textrm{eff}/d\phi^2$ plays the role of an effective mass squared. Outside the massive body, the solution to Eq.~(\ref{eq:Newtonian}) is 
\begin{equation}
\delta \phi = \left. \frac{d \ln A}{d\phi} \right|_{\phi_0} \frac{f(M,R)}{r} e^{-m_\textrm{eff}(\phi_0) r},
\end{equation}
where $f(M,R)$ is a function of the body's mass and radius, determined by matching the exterior and interior solutions. In order to suppress the fifth force mediated by the scalar field in the vicinity of the massive body, one either needs the value of $f(M,R)$ to be small, the effective mass to be large, or the coupling to matter to be small \cite{Burrage2018}. Chameleon theories rely on a combination of an environment-dependent effective mass with a ``thin-shell'' effect, whereby $f(M,R)$ only receives contribution of a thin shell of matter close to the stellar radius, and is thus suppressed. Symmetron and dilaton models rely on the environmental dependence of the conformal coupling.

\subsection{Models}

In this work we focus on two examples of theories with screening mechanisms: the chameleon and environmentally-dependent dilaton models.

In the case of chameleons, we consider a power-law (runaway) potential and an exponential conformal coupling:
\begin{equation} \label{eq:chameleon}
V(\phi) = \mu^{n+4} \phi^{-n}, \qquad
A(\phi) = \exp (\phi/M_c).
\end{equation}
The parameter $M_c$ sets the strength of the coupling between the scalar field and matter, while $\mu$ determines the contribution of the scalar field to the energy density of the universe. For $n \sim 1$ and $M_c \sim M_\textrm{Pl}$, tests of the equivalence principle require that $\mu \lesssim 10^{-3} \textrm{eV} \approx 10^{-30} M_\textrm{Pl}$ \cite{Khoury2004a}.
Since the introduction of the chameleon model, much work has been done to constrain its parameter space with multiple probes; see e.g.~Refs.~\cite{Mota2007,Burrage2016a,Burrage2018} for a discussion on current bounds.

For the environmentally dependent dilaton, we consider an exponentially runaway potential and a quadratic coupling function \cite{Brax2010},
\begin{equation}\label{eq:dilaton}
V(\phi) = V_0 A(\phi)^4 e^{-\Phi(\phi)}, \quad
A(\phi) = 1 + \frac{A_2}{2} \Phi (\phi)^2,
\end{equation}
where $\Phi (\phi)$ is determined implicitly through
\begin{equation} \label{eq:fieldtransf}
d\phi = \sqrt{2} M_\textrm{Pl} \lambda^{-1} 
\sqrt{1 + 3 \lambda^2 \left(\frac{d \ln A}{d\Phi}  \right)^2} d\Phi.
\end{equation}
The quadratic coupling in Eq.~(\ref{eq:dilaton}), valid around its minimum, i.e., for $\Phi \approx 0$, is inspired by the Damour-Polyakov effect \cite{Damour1994}. The field transformation (\ref{eq:fieldtransf}) translates between the original string-inspired effective action with a dilatonic field to the representation (\ref{eq:generalaction}) used in this work. For a local suppression of the coupling to occur, $A_2 \gg 1$, while the scale of $V_0$ is typically set by the dark energy density, and $\lambda \gtrsim O(1)$ \cite{Brax2010}.


It is worth emphasizing that the purpose of the present work is not to investigate whether and which additional constraints could be set to the parameter space of the models above, but rather to point out to new features arising in the environment of highly compact neutron stars. Therefore, although we try to push as close to realistic values for parameters as numerically feasible, we will not strictly adhere to this viable range.
We also ignore the various subtleties that can affect these models, such as the role of quantum corrections \cite{Upadhye2012,Erickcek2013,Erickcek2014}.

\subsection{Equation of state} \label{sec:eos}

Neutron stars are typically well described by a perfect fluid energy-momentum tensor,
\begin{equation} \label{eq:perfectfluid}
T^{\mu\nu} = \epsilon u^\mu u^\nu + p P^{\mu\nu},
\end{equation}
together with a one parameter equation of state. Here $u^\mu$ denotes the four-velocity of fluid elements, $P^{\mu\nu}\equiv g^{\mu\nu} + u^\mu u^\nu$, and $\epsilon$ and $p$ are the energy density and pressure measured in the fluid's rest frame. Alternatively, we can define the Jordan-frame energy-momentum tensor in analogy with Eq.~(\ref{eq:perfectfluid}), in terms of the four-velocity field $\tilde{u}^\mu$, energy density $\tilde{\epsilon}$, and pressure $\tilde{p}$, which satisfy $\tilde{u}^\mu = A(\phi)^{-1} u^\mu$, $\tilde{p} = A(\phi)^{-4} p$, and $\tilde{\epsilon} = A(\phi)^{-4} \epsilon$. 

The equation of state condenses all the complex microphysics of the NS interior in a relation, say, between pressure and rest-mass density, $\tilde{p} = \tilde{p}(\tilde{\rho})$. We choose to specify the EoS in terms of Jordan-frame quantities since in this frame the usual thermodynamic relation for energy conservation holds: $d (\tilde{\epsilon}/\tilde{\rho}) = - \tilde{p} d(1/\tilde{\rho})$.

In this work we will employ the piecewise-polytropic parametrization of Ref.~\cite{Read2009} for the nuclear EoS.
It uses a fixed crust model, based on the SLy EoS, which is continuously connected to three polytropic phases, in which $\tilde{p} = K_i \tilde{\rho}^{\gamma_i}$, with polytropic exponents $\gamma_1$, $\gamma_2$, and $\gamma_3$. The dividing density between the first and second phases is $\rho_1 = 10^{14.7}$g/cm$^{3}$, while the division between the second and third phases occurs at $\rho_2 = 10^{15.0}$g/cm$^3$. A fourth parameter, the pressure $p_1 = \tilde{p}(\rho_1)$ at $\tilde{\rho} = \rho_1$, can be used to determine the crust-core separation. This parametrization has been shown to reproduce the main NS bulk observables predicted by theoretical models based on a wide range of physical assumptions, roughly within a percent accuracy \cite{Read2009}.

\begin{figure}[bt]
\includegraphics[width=8.5cm]{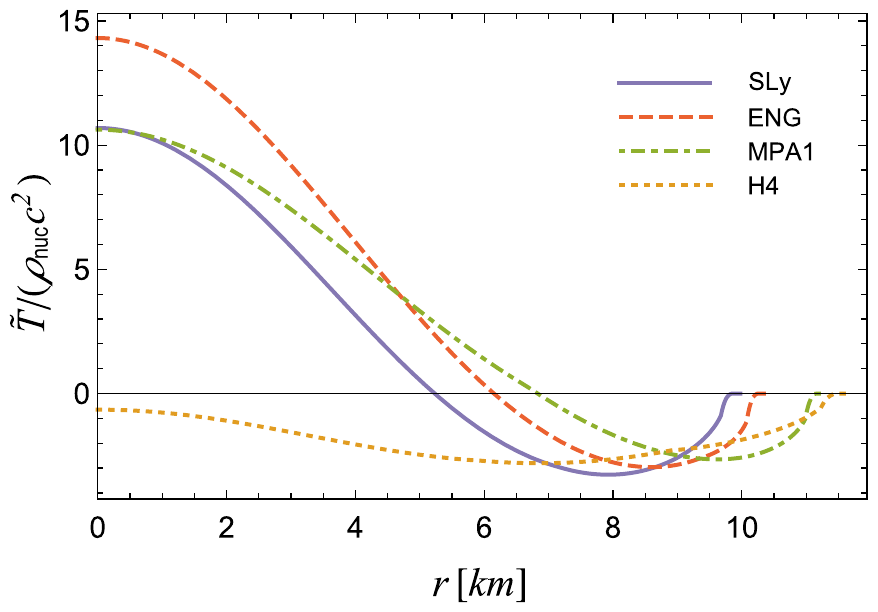}
\caption{Radial profile of $\tilde{T} = 3\tilde{p} - \tilde{\epsilon}$ for the most massive stars allowed by the SLy, ENG, MPA1, and H4 EoS in GR.}
\label{fig:trace}
\end{figure}

In what follows, we will use a piecewise-polytropic approximation to the SLy, ENG, MPA1, and H4 equations of state (see Table III of Ref.~\cite{Read2009} for the best fit values of $\{p_1, \gamma_1, \gamma_2, \gamma_3\}$ for these EoS). In Fig.~\ref{fig:trace} we represent the radial profile of the trace of the energy-momentum tensor inside the most massive star predicted by each of these models (obtained by solving the general relativistic equations of hydrostatic equilibrium). We see that a pressure-dominated phase (with $\tilde{T}>0$) can indeed occur in the core of such stars. Of the four EoS employed in this work, H4 is the only one which does not allow a pressure-dominated phase inside any stable configuration, and was included for the sake of comparison. 

Outside the star, we assume that space is filled with a cosmological fluid, such that $\tilde{p} = - \tilde{\epsilon}$. Note that, in order for the scalar field to settle to a constant value far away from the star ($\phi \to \phi_\infty$ for $r \gg R$), the effective potential must obey $dV_\textrm{eff}/d\phi |_{\phi_\infty} = 0$. In the case of chameleon models characterized by a runaway potential with no extrema, assuming a background matter density outside the star is actually required for the effective potential to have a minimum in this region [cf.~Eq.~(\ref{eq:field2})]. The reason for choosing the cosmological equation of state $\tilde{p} = - \tilde{\epsilon}$ is twofold. First, this allows us to obtain self-consistent solutions for the equilibrium equations [since $d\tilde{p}/dr = 0$ in Eq.~(\ref{eq:eqpb}) below]. Second, and more important, it ensures that the spacetime is asymptotically Schwarzschild-de Sitter, which allows for clearer boundary conditions to be imposed on our metric functions (see Sec.~\ref{sec:equations}). 

When dealing with the chameleon model, we set the background energy density to be $\tilde{\epsilon}_\infty = 3.9 \times 10^{-4} \rho_\textrm{nuc} c^2$ for computational reasons, where $\rho_\textrm{nuc} \equiv 1.66 \times 10^{14}$ g/cm$^{3}$ is a reference density of the order of the nuclear saturation density, which will be used throughout. On the other hand, there is no technical need to introduce a background matter density outside the star in the dilaton model, since the potential dependence on $A(\phi)$ already ensures it has a minimum. Therefore, in this case we set $\tilde{\epsilon}_\infty = 0$ for simplicity.

\section{Equilibrium and perturbation equations} \label{sec:equations}

\subsection{Structure equations}

We begin by considering static, spherically symmetric NS configurations, with a perfect fluid energy-momentum tensor. The spacetime can be described by the line element
\begin{equation}\label{eq:lineelement1}
ds^2 = - e^{2 \nu (r)} dt^2 + e^{2 \lambda(r)} dr^2 + r^2 (d\theta^2 + \sin^2\theta d\varphi^2).
\end{equation}
Under the assumptions of staticity and spherical symmetry, and defining the mass aspect function $m(r) \equiv (r/2) (1-e^{-2\lambda})$, the field equations (\ref{eq:field1}) and (\ref{eq:field2}) can be written as
\begin{align}
&\frac{dm}{dr} = 4\pi r^2 \left[ A^4 \tilde{\epsilon} + \frac{1}{2} e^{-2\lambda} \psi^2 + V \right],
\label{eq:eqmb}
\\
&\frac{d\nu}{d r} = r e^{2\lambda} \left[ \frac{m}{r^3} + 4\pi A^4 \tilde{p} + 2\pi e^{-2\lambda} \psi^2 - 4\pi V \right], \label{eq:eqnub}
\\
&\frac{d\tilde{p}}{d r} = - (\tilde{p}+\tilde{\epsilon}) \frac{d}{dr} \left( \nu + \ln A \right),
\label{eq:eqpb}
\\
& \frac{d\phi}{dr} = \psi\\
&\frac{d}{dr}\! \left( r^2 e^{\nu - \lambda} \psi \right) = r^2 e^{\nu + \lambda} \left[ \frac{d V}{d \phi} -  A^3 \frac{d A}{d\phi} (3\tilde{p} - \tilde{\epsilon})\right].
\label{eq:eqphib}
\end{align}
When supplemented by a choice of model functions $A(\phi)$ and $V(\phi)$ and of EoS, these equations can be integrated numerically by standard methods. The boundary conditions are the following. We require the solution to be analytic around $r=0$, which sets $m(0)=0$. The stellar radius is determined through the condition $\tilde{p}(R) = \tilde{p}_\infty = - \tilde{\epsilon}_\infty$; the need to assume an atmosphere outside the star in the chameleon model was discussed in Sec.~\ref{sec:eos}, as well as the choice of a cosmological fluid. In practice, $\tilde{p}(R) = 0$ gives a good estimate for the stellar radius, since $\tilde{\epsilon}_\infty$ is small. Far away from the star, we require the scalar field to asymptote to a constant, $\phi_\infty$. The value of $\phi_\infty$ can be obtained from Eq.~(\ref{eq:eqphib}), and corresponds to a minimum of the effective potential, i.e., a solution of
\begin{equation}
\left. \frac{d V}{d \phi} \right|_{\phi_\infty} = A^3 (\phi_\infty) \left. \frac{d A}{d\phi}\right|_{\phi_\infty} (3\tilde{p}_\infty - \tilde{\epsilon}_\infty).
\end{equation}
In order to enforce the latter boundary condition, we implemented a shooting algorithm, whereby the central value of the scalar field is adjusted until the proper asymptotic behavior is verified outside the star. It might be worth mentioning that, due to a high sensitivity to the initial conditions, we had to resort to more than double machine precision to achieve convergence in some cases.

Note that, since Eqs.~(\ref{eq:eqmb})-(\ref{eq:eqphib}) do not depend on $\nu(r)$ except through its radial derivative, the boundary condition for this metric function is irrelevant as far as equilibrium stellar properties are concerned. However, the star's oscillation frequencies are sensitive to the normalization of this metric function. We fix it by imposing that far away from the star the spacetime should become Schwarzschild-de Sitter, 
\begin{equation} \label{eq:SdS}
ds^2 = - f(r) dt^2 + f(r)^{-1} dr^2 + r^2 (d\theta^2 + \sin^2 \theta d\varphi^2),
\end{equation}
with $f(r) = 1 - 2a/r - br^2$. In practice, we require that $\nu(r) \to (1/2) \ln [1-2m(r)/r] $ for $r\gg R$.
Additionally, the total mass is identified with the parameter $a$ when matching the numerical solution of Eqs.~(\ref{eq:eqmb})-(\ref{eq:eqphib}) to the asymptotic form (\ref{eq:SdS}) far away from the star. Explicitly, we have:
\begin{equation}
M \approx m(r) - \frac{4\pi}{3} r^3 \left[ A(\phi_\infty)^4 \tilde{\epsilon}_\infty + V(\phi_\infty) \right],  \quad r\gg R.
\end{equation}
Numerically, $M$ differs only slightly from $m(R)$, which was used in previous works as an estimate for the stellar mass. 

Finally, the total baryon mass of the star can be computed from 
\begin{equation}
M_b = \int_0^R 4\pi r^2 \tilde{\rho} A(\phi)^3 (1-2m/r)^{-1/2} dr.
\end{equation}

\subsection{Radial perturbations} \label{sec:perturbations}

In this subsection, we derive the full set of equations describing linear, adiabatic, radial perturbations of equilibrium solutions in models described by Eq.~(\ref{eq:generalaction}). These equations generalize those presented in Ref.~\cite{Mendes2018} to a nonzero potential, and are a subcase of the tensor-multi-scalar perturbation equations recently discussed in Ref.~\cite{Doneva2020}. Radial stellar oscillations have also been considered in chameleonlike theories in the Newtonian context \cite{Sakstein2013}.

Here we show how all perturbed quantities can be written in terms of the Lagrangian displacement $\xi(t,r)$ and the scalar field perturbation $\delta \phi (t,r)$, which obey a set of coupled second order differential equations. 
In particular, these equations govern the stability of a relativistic star to gravitational collapse. We discuss the relevant boundary conditions for unstable modes, and the numerical procedure we employ to search for such solutions. 
The case of stable radial perturbations is more involved due to the presence of scalar radiation, which implicates boundary conditions that are more subtle to implement numerically. This case was recently analyzed in Ref.~\cite{Mendes2018} for theories with a trivial potential, and will be investigated for models with screening mechanisms in a future work.

\subsubsection{Perturbation equations} 

Since spherical symmetry is retained in the perturbed configuration, the line element can be conveniently written as
\begin{equation}
ds^2 = - e^{2 \nu(t,r)} dt^2 + e^{2\lambda(t,r)} + r^2 (d\theta^2 + \sin^2 \theta d\varphi^2),
\end{equation}
where $\nu(t,r) = \nu_0 (r) + \delta \nu (t,r)$ and $\lambda(t,r) = \lambda_0 (r) + \delta \lambda (t,r)$, with $\nu_0(r)$ and $\lambda_0(r) = - (1/2) \ln [1-2m(r)/r]$ denoting the background quantities obeying Eqs.~(\ref{eq:eqmb}) and (\ref{eq:eqnub}). Similarly, we write the perturbed scalar field as $\phi(t,r) = \phi_0(r) + \delta \phi(t,r)$, with $\psi_0(r) = d\phi_0/dr$ obeying Eq.~(\ref{eq:eqphib}).

The perturbed (Einstein-frame) fluid four-velocity is given by 
\begin{equation}
u^\mu (t,r) = e^{-\nu_0} (1-\delta \nu, d\xi/dt, 0, 0),
\end{equation}
where $\xi(t,r)$ is the radial Lagrangian displacement of a given fluid element. The perturbed pressure and energy density are written as $p(t,r) = p_0(r) + \delta p(t,r)$ and $\epsilon(t,r) = \epsilon_0(r) + \delta \epsilon(t,r)$. The corresponding Jordan-frame quantities are
\begin{align*}
\delta \tilde{u}^\mu &= A(\phi_0)^{-1} (\delta u^\mu - \alpha_0 u^\mu \delta \phi), \\
\delta \tilde{p} &= A(\phi_0)^{-4} (\delta p - 4 \alpha_0 p_0 \delta \phi ), \\
\delta \tilde{\epsilon} &= A(\phi_0)^{-4} (\delta \epsilon - 4  \alpha_0 \epsilon_0 \delta \phi ),
\end{align*}
where we used the shorthand
\begin{equation}
\alpha_0 \equiv \left.\frac{d \ln A}{d\phi}\right|_{\phi_0}.
\end{equation}

Therefore, the perturbed configuration is fully characterized by six functions of $t$ and $r$, namely, $\delta \nu$, $\delta \lambda$, $\delta \phi$, $\xi$, $\delta \tilde{p}$, and $\delta \tilde{\epsilon}$.
For adiabatic perturbations, and assuming that the perturbed fluid has the same EoS as the unperturbed configuration, pressure and energy density perturbations can be written in terms of the mass-density perturbation $\delta \tilde{\rho}$ as
\begin{equation} \label{eq:deltap}
\delta  \tilde{p} = \frac{\Gamma_1 \tilde{p}_0}{\tilde{\rho}_0} \delta \tilde{\rho}, \qquad
\delta \tilde{\epsilon} = \frac{\tilde{\epsilon}_0 + \tilde{p}_0}{\tilde{\rho}_0} \delta \tilde{\rho},
\end{equation}
where
$$
\Gamma_1 = \left(\frac{\partial \ln \tilde{p}}{\partial \ln \tilde{\rho}} \right)_s
$$
is the adiabatic index (defined at constant entropy $s$).

The perturbed equation for rest-mass conservation, $\delta [\tilde{\nabla}_\mu (\tilde{\rho} \tilde{u}^\mu)] = 0$, can be directly integrated to yield
\begin{equation} \label{eq:deltarho}
\frac{\delta \tilde{\rho}}{\tilde{\rho}_0} = - \delta \lambda - 3 \alpha_0 \delta \phi - \frac{\partial \xi}{\partial r} - \xi \left( \frac{2}{r} + \frac{d\lambda_0}{dr} + 3 \alpha_0 \psi_0 + \frac{1}{\tilde{\rho}_0} \frac{d \tilde{\rho}_0}{dr} \right),
\end{equation}
which relates the mass density perturbation to $\delta \lambda$, $\delta \phi$, and $\xi$. Similarly, by perturbing Eq.~(\ref{eq:field1}), one obtains, after some manipulation,
\begin{equation} \label{eq:deltalambda}
\delta \lambda = 4 \pi r \left[ \psi_0 \delta \phi - A(\phi_0)^4 e^{2 \lambda_0} (\tilde{\epsilon}_0 + \tilde{p}_0) \xi \right]
\end{equation}
($tr$ component) and
\begin{align} \label{eq:ddnu}
\frac{\partial \delta\nu}{\partial r} &=  \delta \lambda e^{2\lambda_0} \left[ \frac{1}{r} + 8\pi r p_0 - 8\pi r V(\phi_0) \right] - 4\pi r \psi_0 \frac{\partial \delta \phi}{\partial r} \nonumber \\
&+ 4\pi r e^{2\lambda_0} \delta p
\end{align}
($rr$ component). In Eq.~(\ref{eq:ddnu}) the Einstein-frame background pressure and pressure perturbation were used to abbreviate the expression. 

Next, perturbation of Eq.~(\ref{eq:field2}) yields
\begin{align} \label{eq:deltaphi}
0 &= e^{2(\lambda_0 - \nu_0)} \frac{\partial^2 \delta \phi}{\partial t^2} - \frac{\partial^2 \delta \phi}{\partial r^2} + \frac{\partial \delta \phi}{\partial r} \left[ -\frac{2}{r} + \frac{d\lambda_0}{dr} - \frac{d\nu_0}{dr} \right] \nonumber \\
& + e^{2\lambda_0} \delta \phi \left[ \left.\frac{d^2 V}{d\phi^2} \right|_{\phi_0} - A(\phi_0)^4 (3 \tilde{p}_0 - \tilde{\epsilon}_0) ( 4 \alpha_0^2 + \beta_0 )\right] \nonumber \\
& + \psi_0 \frac{\partial}{\partial r} (\delta \lambda - \delta \nu) + 2 \delta \lambda \left[ \psi_0 \left( \frac{2}{r} - \frac{d \lambda_0}{dr} + \frac{d \nu_0}{dr} \right) + \frac{d\psi_0}{dr} \right] \nonumber \\
& - A(\phi_0)^4  e^{2\lambda_0} \alpha_0 (3 \delta \tilde{p} - \delta \tilde{\epsilon}),
\end{align}
where the shorthand 
$$
\beta_0 \equiv \left.\frac{d^2 \ln A}{d\phi^2}\right|_{\phi_0}
$$
was used.
Finally, perturbation of the equations of motion (\ref{eq:eqmotion}) leads to
\begin{align}\label{eq:xieq}
0 & = e^{2(\lambda_0 - \nu_0)} \frac{\partial^2 \xi}{\partial t^2} + \alpha_0 \frac{\partial \delta \phi}{\partial r} + \beta_0 \psi_0 \delta \phi + \frac{1}{(\tilde{p}_0 + \tilde{\epsilon_0})} \frac{\partial \delta \tilde{p}}{\partial r} \nonumber \\
& + \frac{1}{(\tilde{p}_0 + \tilde{\epsilon_0})^2} \frac{d \tilde{p}_0}{dr} (\delta \tilde{p} + \delta \tilde{\epsilon}) + \frac{\partial \delta \nu}{\partial r}.
\end{align}

Substituting Eqs.~(\ref{eq:deltap}), (\ref{eq:deltarho}), (\ref{eq:deltalambda}), and (\ref{eq:ddnu}) into Eqs.~(\ref{eq:deltaphi}) and (\ref{eq:xieq}), one obtains two master equations for $\xi$ and $\delta \phi$, comprising a system of coupled homogeneous second order partial differential equations, with coefficients that depend solely on background quantities. We will not write these equations explicitly here, since their full form is not particularly illuminating, and can be obtained straightforwardly by the procedure described above.

Next, we assume a harmonic time dependence for the perturbation variables $\xi$ and $\delta \phi$:
\begin{equation}\label{eq:harmonict}
\xi (t,r) = \xi (r) e^{i\omega t}, \qquad
\delta \phi (t,r) = \delta \phi (r) e^{i \omega t},
\end{equation}
with $\omega \in \mathbb{C}$. Stable (unstable) modes are characterized by $\Im (\omega) > 0$ ($\Im (\omega) < 0$). With the ansatze (\ref{eq:harmonict}), and defining the vector function ${\bf x} (r) = (\xi, \xi', \delta \phi, \delta \phi')^T$ (with a prime denoting a radial derivative), our master equations assume the form
\begin{equation} \label{eq:diffall}
\frac{d {\bf x} (r)}{dr} = {\bf M}(r) {\bf x}(r),
\end{equation}
where ${\bf M}(r)$ is a $4\times 4$ matrix function of background quantities alone.

\begin{figure*}[thb]
\includegraphics[width=15cm]{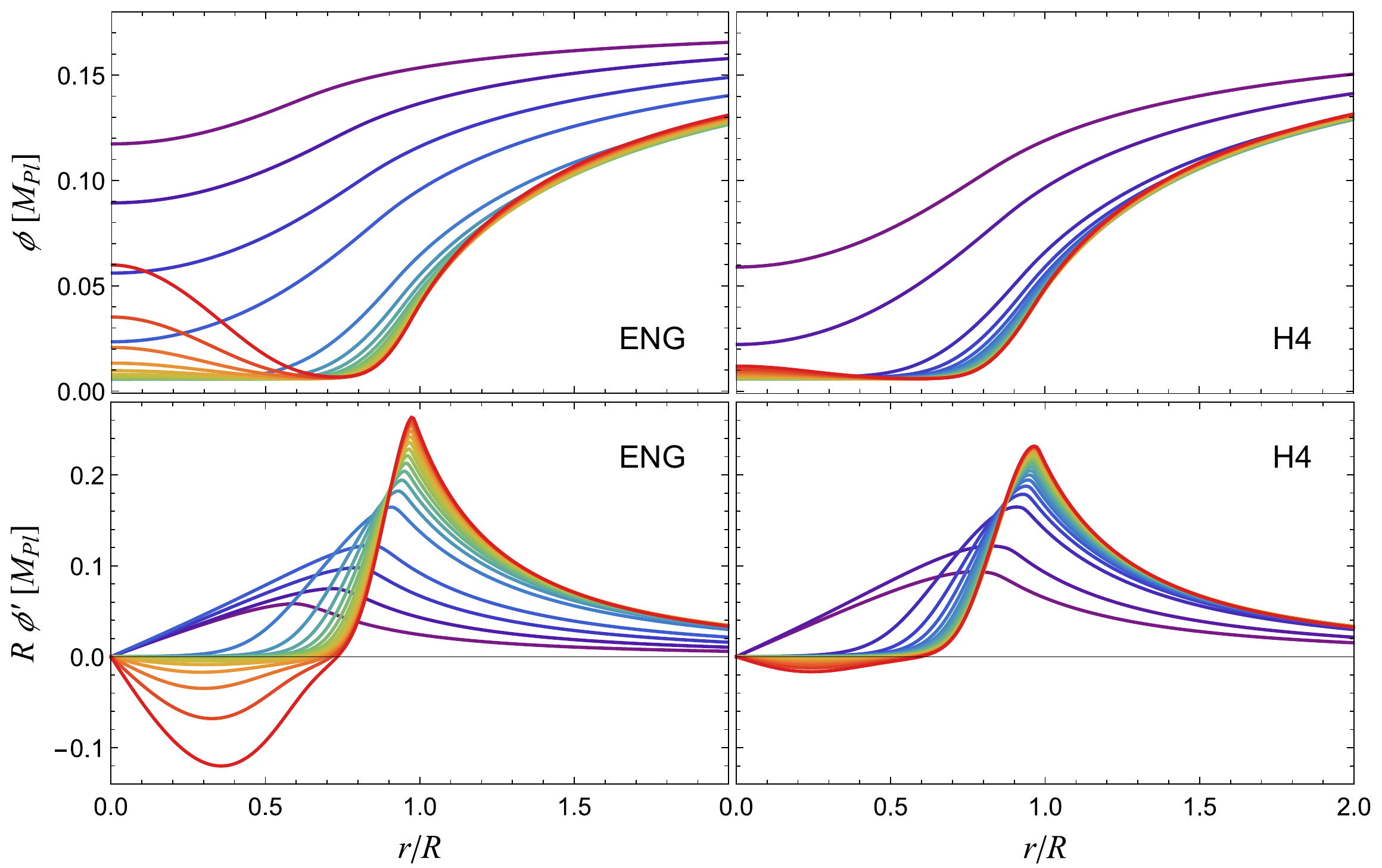}
\caption{\textit{Top panels}: Scalar field as a function of the radial coordinate, for the ENG and H4 EoS. \textit{Bottom panels}: Scalar field gradient as a function of the radial coordinate, for the ENG and H4 EoS. Colors from violet to red indicate increasing central densities (all of which yield stable stars): from $2.2\rho_\textrm{nuc}$ to $8.6\rho_\textrm{nuc}$ for the ENG EoS and from $2.2\rho_\textrm{nuc}$ to $10.6\rho_\textrm{nuc}$ for the H4 EoS, both in steps of $0.4 \rho_\textrm{nuc}$. Here we consider the chameleon model of Eq.~(\ref{eq:chameleon}) with $n = 1$, $M_c = M_\textrm{Pl}$, and $\mu = 7.2 \times 10^{-17} M_\textrm{Pl}$. As the central density increases and the star becomes more compact, we see a partial unscreening of the scalar field in the core of NSs described by the ENG EoS.}
\label{fig:fieldprofile}
\end{figure*}

\begin{figure*}[htb]
\includegraphics[width=17cm]{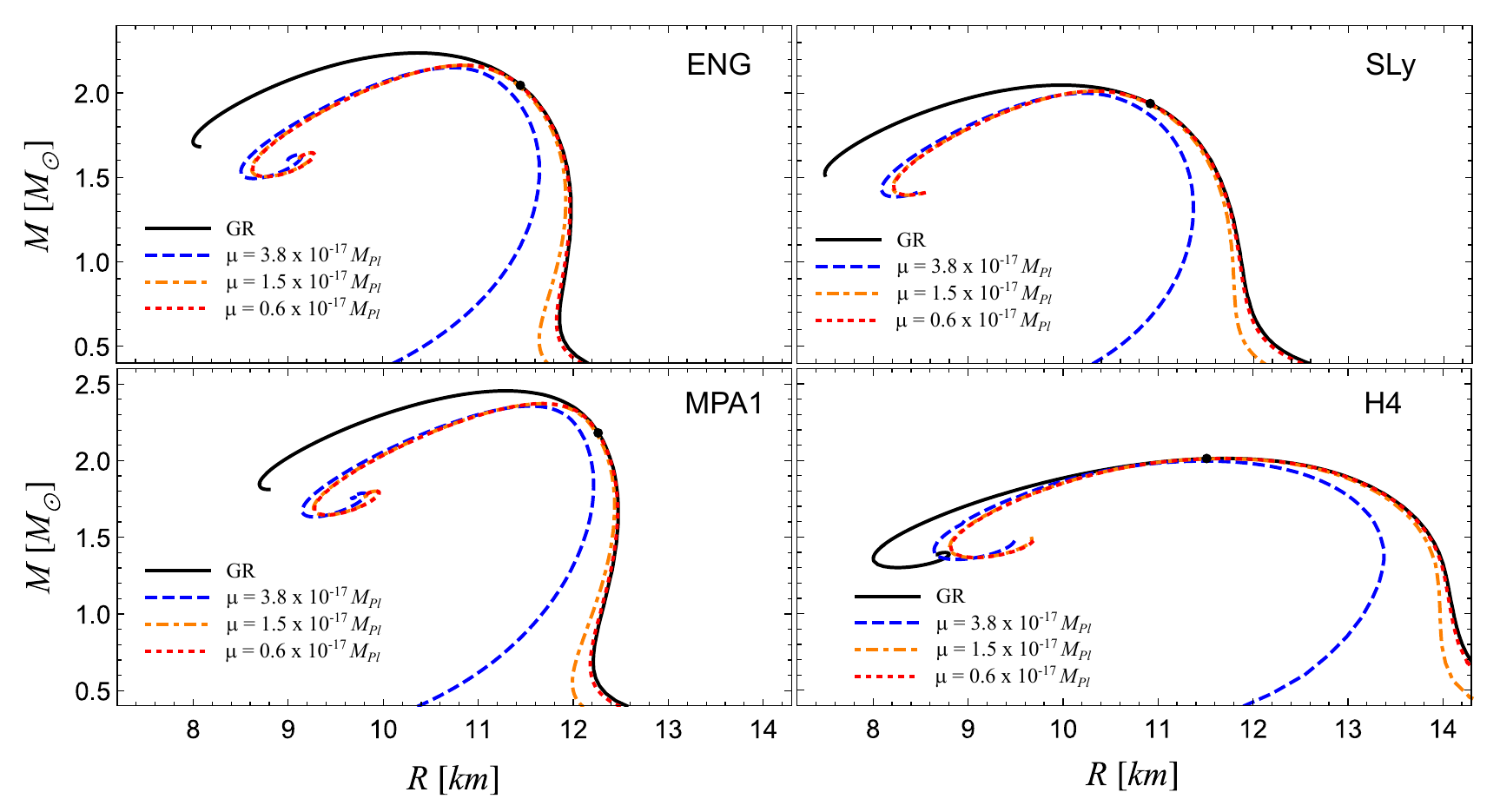}
\caption{Mass-radius curves for the chameleon model (\ref{eq:chameleon}), with $n=1$, $M_c=M_\textrm{Pl}$, and three values of $\mu$. The prediction from GR is shown in solid black for comparison. Black dots indicate configurations along the GR curve starting from which $\tilde{T}>0$ in a region of the stellar interior. For the H4 EoS, a pressure-dominated phase only occurs for dynamically unstable stars (i.e.~after the turning point).}
\label{fig:massradius}
\end{figure*}

\subsubsection{Boundary conditions and integration procedure}

The relevant boundary and junction conditions for $\xi(r)$ and $\delta \phi(r)$ are the following.
\begin{itemize}

\item[(i)] Regularity at $r = 0$, which requires that $\xi(0) = 0$ and $\delta \phi'(0) = 0$. 

\item[(ii)] Jump in $\xi'(r)$ at transition between polytropic phases. In Eq.~(\ref{eq:xieq}), the term proportional to $\partial \delta \tilde{p}/\partial r$ depends on $\Gamma_1'(r)$, which is the derivative of a piecewise constant function, and therefore given by a sum of Dirac delta functions at the radii $r_i$ corresponding to transitions between the various polytropic phases. This implies that $\xi'(r)$ will not be continuous, but will experience a jump at each transition radius, according to
\begin{align} \label{eq:jumpdxi}
\Delta_i(\Gamma_1 \zeta') & = \Delta_i \Gamma_1 
\{ A(\phi_0)^4 r^2 e^{-\nu_0} [ \xi \psi_0 (\alpha_0 - 4\pi r \psi_0) \nonumber \\
& \left.- \delta \phi (3 \alpha_0 + 4\pi r \psi_0)]\}\right|_{r=r_i},
\end{align}
where $\zeta \equiv e^{-\nu_0} A(\phi_0)^4 r^2 \xi$ and $\Delta_i Q \equiv \lim_{\epsilon \to 0} [Q(r_i + \epsilon) - Q(r_i - \epsilon)]$ denotes the discontinuity of a quantity $Q$ across $r = r_i$.

\item[(iii)] Regularity at the stellar surface. The perturbed stellar surface, located at $R_\textrm{new} = R + \xi(R)$, must satisfy $\tilde{p}(R_\textrm{new})=0$, which implies
\begin{equation} \label{eq:pRnew}
\tilde{p}'_0(R) \xi(R) + \delta \tilde{p} (R) = 0.
\end{equation}
As in GR, the left-hand side automatically vanishes as long as $\xi$, $\delta \phi$, and their derivatives are finite at $r=R$. Therefore, it suffices to impose regularity of the perturbed variables at $r = R$.

By examining the full form of the differential equations (\ref{eq:diffall}), we find potentially diverging terms as $r \to R$, proportional to $\tilde{\epsilon}_0/\tilde{p}_0 \sim \tilde{\rho}_0^{1 - \gamma_R}$ and $\tilde{\epsilon}_0^2/\tilde{p}_0 \sim \tilde{\rho}_0^{2 - \gamma_R}$, where $\gamma_R$ is the polytropic exponent at the outermost polytropic layer. Assuming $\gamma_R<2$, as will be the case for all EoS considered in this work, we demand that the coefficient of the term $\tilde{\epsilon}_0/\tilde{p}_0$ vanishes at $r=R$. The resulting condition has the form
\begin{equation} \label{eq:bcR}
{\bf F}(R)^T {\bf x}(R) = 0,
\end{equation}
where ${\bf F}(R)$ is a vector function of background quantities.

As discussed in Sec.~\ref{sec:eos}, it is technically necessary to postulate a nonzero energy density outside the star in the chameleon model --- but not so for the environmentally-dependent dilaton. 
It should be noted that we are neglecting perturbations of the atmosphere in which the star may be immersed; additionally, Eq.~(\ref{eq:pRnew}) is not exact when there is fluid outside the star. However, since the energy density $\tilde{\epsilon}_\infty$ is assumed to be very low, the numerical errors incurred in assuming the validity of Eq.~(\ref{eq:pRnew}) will also be small.

\item[(iv)] Finally, we demand that $\delta \phi (r) \to 0$ far away from the star, which is the appropriate boundary condition for unstable modes, i.e. when $\omega^2 = -\Omega^2 <0$. In the case of stable perturbations, this condition would have to be substituted by an outgoing boundary condition at the cosmological horizon.

\end{itemize}

With the boundary conditions specified above, Eq.~(\ref{eq:diffall}) can be solved numerically by standard methods. We conclude this subsection with a brief outline of the integration procedure. First, note that any solution of Eq.~(\ref{eq:diffall}) can be written as a set of four linearly independent solutions ${\bf x}_i$ ($i \in \{1,...,4\}$):
$$
{\bf x} (r) = c_1 {\bf x}_1(r) + c_2 {\bf x}_2(r) + c_3 {\bf x}_3(r) + c_4 {\bf x}_4(r),
$$
where $c_i$ are constants. If we choose four linearly independent vectors ${\bf x}_i^\textrm{in}(0)$ such as ${\bf x}_1^\textrm{in}(0) = (0,1,1,0)^T$, ${\bf x}_2^\textrm{in}(0) = (0,1,-1,0)^T$, ${\bf x}_3^\textrm{in}(0) = (1,1,1,0)^T$, and ${\bf x}_4^\textrm{in}(0) = (0,1,1,1)^T$, then the boundary conditions (i) imply that $c_3 = c_4 = 0$. The unique solution satisfying (i) can be written as ${\bf x}^\textrm{in} (r) = c_1 {\bf x}_1^\textrm{in}(r) + c_2 {\bf x}_2^\textrm{in}(r)$, where ${\bf x}_1^\textrm{in}(r)$ and ${\bf x}_2^\textrm{in}(r)$ are obtained by integrating Eq.~(\ref{eq:diffall}) from $r=0$ with the initial conditions above, and taking care of the derivative jumps implicated by (ii). Similarly, we can find three linearly independent vectors (say ${\bf x}_1^\textrm{out}(R)$, ${\bf x}_2^\textrm{out}(R)$, and ${\bf x}_3^\textrm{out}(R)$) that obey the boundary condition (\ref{eq:bcR}) at $r = R$. The unique solution satisfying (iii) is then written as ${\bf x}^\textrm{out} (r) = a_1 {\bf x}^\textrm{out}_1(r) + a_2 {\bf x}_2^\textrm{out}(r) + a_3 {\bf x}_3^\textrm{out}(r)$, with ${\bf x}_i^\textrm{out}(r)$ obtained from the numerical integration of Eq.~(\ref{eq:diffall}) from $r=R$ inwards, again taking care of the jumps implied by (ii). By matching the solutions ${\bf x}^\textrm{in} (r)$ and ${\bf x}^\textrm{out}(r)$ at some intermediate radius, say $r=R/2$, one obtains four algebraic conditions that must be satisfied by the five arbitrary constants $c_1$, $c_2$, $a_1$, $a_2$, and $a_3$. Since the overall normalization of ${\bf x}(r)$ is arbitrary, it can be fixed, e.g., by requiring that $\delta\phi(0)=1$, which implies $c_2 - c_1 = 1$. This closes the system and enables the computation of the interior solution unequivocally, given a value of $\omega^2$. Finally, with the conditions for $\delta \phi(R)$ and $\delta \phi'(R)$ coming from the solution to the inner problem, we evolve the scalar field equation outside the star, and implement a shooting procedure to find the values of $\omega^2 = -\Omega^2 <0$ such that condition (iv) is verified asymptotically.

\section{Results} \label{sec:results}

\subsection{Chameleon}

We begin our analysis of relativistic stars in the chameleon model, with $V(\phi)$ and $A(\phi)$ given by Eq.~(\ref{eq:chameleon}). We take $n=1$ and $M_c = M_\textrm{Pl}$ throughout, but vary the value of $\mu$. Note that, in order to satisfy equivalence principle constraints, $\mu \lesssim 10^{-30} M_\textrm{Pl}$ \cite{Khoury2004a}. However, reaching such a small value is prohibitive from the numerical standpoint: Indeed, already for $\mu \lesssim 10^{-17} M_\textrm{Pl}$ we find that a high level of fine-tuning is required to obtain solutions with the proper asymptotic behavior. Still, as we discuss below, we believe that our main conclusions would still hold for $\mu$ in the realistic range of values.

Figure \ref{fig:fieldprofile} shows the radial profiles for the scalar field and its gradient, for the ENG and H4 equations of state, and $\mu = 7.2 \times 10^{-17} M_\textrm{Pl}$. In both cases, for low central densities (bluer colors) the scalar field profile is relatively flat and the star is unscreened --- a consequence of the large value of $\mu$ adopted in this plot. As the central density increases, a characteristic thin-shell pattern appears, with the scalar field and its gradient suppressed in the stellar interior. This is at the core of the chameleon screening mechanism \cite{Khoury2004a}. Nonetheless, if the central density is sufficient high (redder colors), stars described by the ENG EoS exhibit a pressure-dominated phase, which re-activates the scalar field in the stellar core. We thus see an amplification of the scalar-mediated fifth force (proportional to $(\phi')^2$) in the stellar interior. However, for all realistic EoS, this pressure-dominated phase does not extend all the way to the stellar surface and the scalar field is again suppressed in the outer layers of the star. As a consequence, the exterior profile for the scalar field and its derivative does not display much difference between the ENG and H4 EoS, the latter of which does not allow a pressure-dominated phase inside any stable star. 

Most interestingly, the reactivation of the scalar field in the stellar core can affect the NS structure, leaving imprints in observable quantities such as their masses and radii. Figure \ref{fig:massradius} shows mass-radius curves in the chameleon model (\ref{eq:chameleon}) with $n = 1$, $M_c = M_\textrm{Pl}$, and three values of $\mu$, for the four EoS considered in this work. As $\mu$ decreases, the contribution from the potential $V(\phi)$ becomes smaller and we see that, for all EoS, the sequences of equilibrium configurations converge to a limiting curve, which tends to GR at low densities, but diverges at high enough densities. This behavior suggests that, if $\mu$ was pushed down to realistic values, masses and radii of NS solutions would not change appreciably with respect to the lowest value of $\mu$ considered in Fig.~\ref{fig:massradius}.

Note that deviation from GR starts as soon as the trace of the energy-momentum tensor becomes positive in some region of the stellar interior, as identified by black dots in the plots. The decrease in the maximum mass, for the lowest value of $\mu$ displayed in Fig.~\ref{fig:massradius}, is of $3.2\%$, $1.7\%$, and $3.4\%$ for the ENG, SLy, and MPA1 equations of state, respectively. This can be contrasted with the decrease of only $0.016\%$ for the H4 EOS, which displays no pressure dominated phase inside stable stars. 

\begin{figure}[htb]
\includegraphics[width=8.5cm]{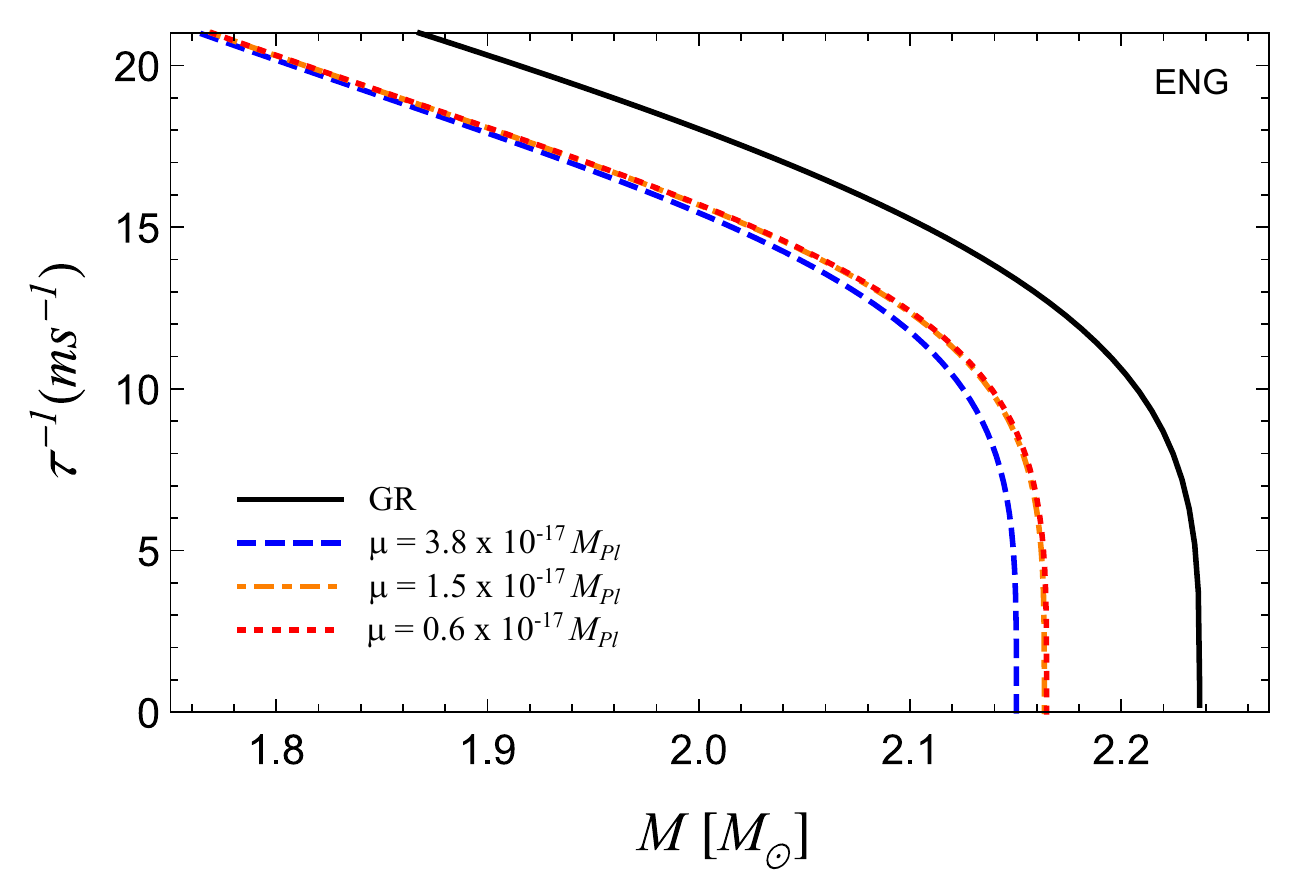}
\caption{Inverse of the instability timescale as a function of total mass for NS solutions described by the ENG EoS, both in GR and in the chameleon model of Eq.~(\ref{eq:chameleon}), with $n = 1$, $M_c = M_\textrm{Pl}$ and the three values of $\mu$ specified in the plot. The maximum mass configuration presents a marginally stable mode, while denser configurations are radially unstable.}
\label{fig:modescham}
\end{figure}

Next, we investigate the radial stability of the equilibrium solutions found previously, following the procedure outlined in Sec.~\ref{sec:perturbations} to search for unstable modes, with time dependence $\exp (\Omega t)$, with $\Omega >0$. 
We find no evidence of unstable modes for configurations lying prior to the turning point in the mass-radius diagrams of Fig.~\ref{fig:massradius}. As in GR, a marginally stable mode, with $\Omega \approx 0$, is found for the maximum-mass solution, and unstable modes are found for denser configurations, with the instability timescale $\tau \equiv \Omega^{-1}$ decreasing as the central density of the solution increases. 
Figure \ref{fig:modescham} displays the inverse of the instability timescale $\tau^{-1}$ as a function of the total mass for the ENG EoS. The corresponding equilibrium solutions are those lying to the left of the turning point in the mass-radius diagram of the upper-left panel of Fig.~\ref{fig:massradius}. For the chameleon model as well as in GR, radial instability sets in at the turning point of sequences of equilibria; the magnitude of $\tau$ is set by the NS dynamical timescale, $\sqrt{R^3/GM}$, which is of the order of milliseconds.

Our analysis makes it clear that stable, partially unscreened NS configurations can exist in the chameleon model, at least for some realistic EoS. This is in contrast with the conclusion arrived at in Refs.~\cite{Babichev2010,Brax2017} by a less rigorous argument, and raises the interesting possibility of using measurements of the most massive, most compact NSs to further constrain these models.

\subsection{Dilaton}

\begin{figure}[th]
\includegraphics[width=8.6cm]{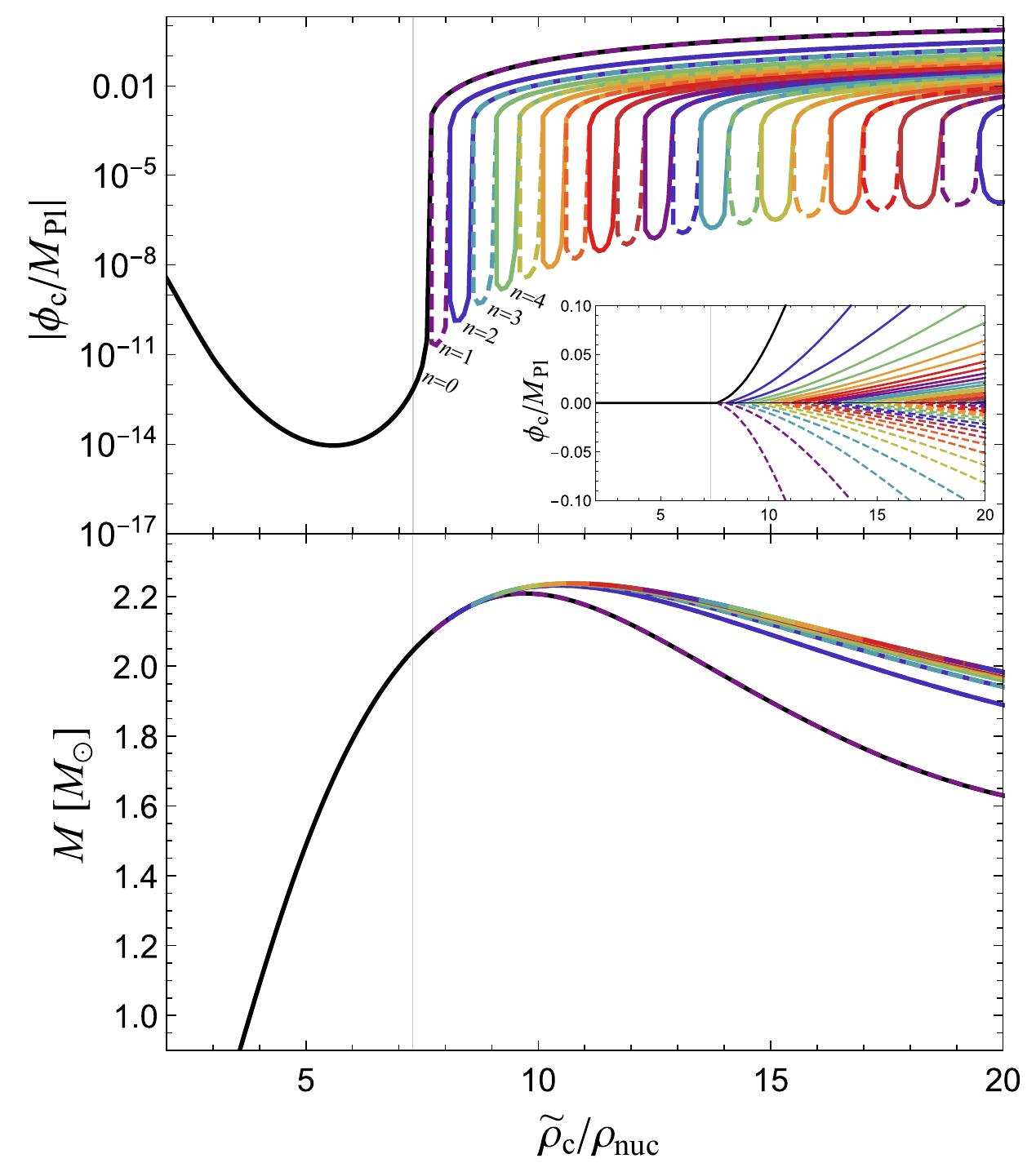}
\caption{Results for the dilaton model (\ref{eq:dilaton}) with  $\lambda = 1$, $A_2 = 1000$, and $V_0 = 3.9 \times 10^{-34} \rho_\textrm{nuc}$, and the ENG EoS. The top panel displays the absolute value of the scalar field at the stellar center, $\phi_c \equiv \phi(r=0)$, as a function of the central density in a log scale, while the inset shows $\phi_c$ in a linear scale, for the same range of densities. Solutions with a different number $n$ of nodes are represented by various colors and we identify the number $n$ for the five first families. Solid (dashed) curves correspond to solutions with a positive (negative) value of $\phi_c$. The critical central density, $\tilde{\rho}_c \approx 7.30 \rho_\textrm{nuc}$, above which $\tilde{T}(r=0)>0$ is displayed as a vertical line. The bottom panel shows the total mass as a function of the central density. Only the first branches are clearly distinguishable, with the high-$n$ solutions accumulating around the GR values.}
\label{fig:dil1}
\end{figure}

\begin{figure}[th]
\includegraphics[width=8.7cm]{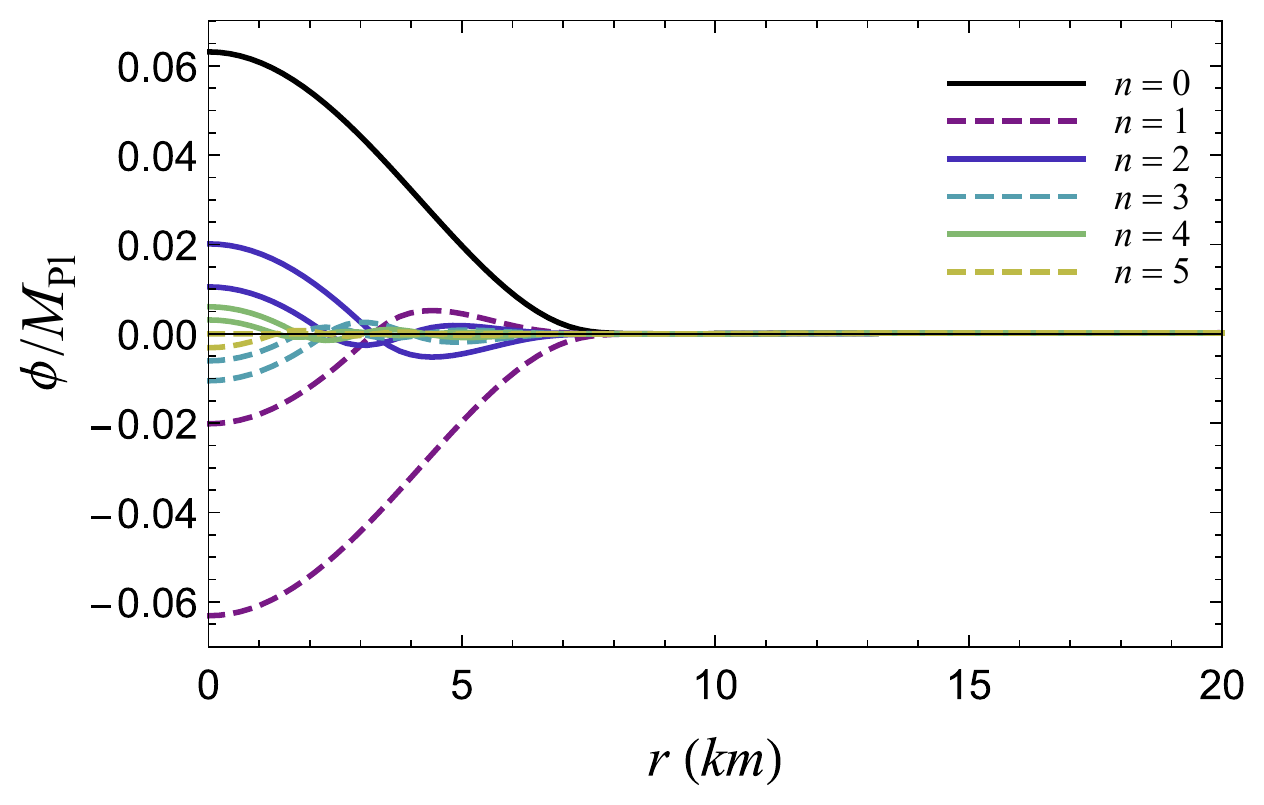}
\caption{Scalar field as a function of the radial coordinate for the eleven equilibrium solutions found with $\tilde{\rho}_c = 10.0 \; \rho_\textrm{nuc}$ in the dilaton model. The equation of state and model parameters are the same as in Fig.~\ref{fig:dil1}. The solutions are assorted by the number $n$ of nodes in the scalar field profile, with the same color coding as in Fig.~\ref{fig:dil1}. All solutions asymptote to $\phi_\infty \approx 3.64 \times 10^{-4} M_\textrm{Pl}$.
The stellar surface is located approximately at 10.5 km.}
\label{fig:fieldprofileDil}
\end{figure}

Let us now turn to the environmentally-dependent dilaton, with $V(\phi)$ and $A(\phi)$ given by Eq.~(\ref{eq:dilaton}). As a representative example, we choose the ENG EoS and fix the model parameters to $\lambda = 1$, $A_2 = 1000$, and $V_0 = 3.9 \times 10^{-34} \rho_\textrm{nuc}$. For this choice of parameters and in the absence of a pressure-dominated phase (i.e., for low central densities), NSs in the dilaton model have very similar structural properties to their GR counterparts, with the scalar field exhibiting a typical thin-shell pattern. However, stars with pressure-dominated cores in the dilaton model can have a widely different behavior, exhibiting the effect of \textit{spontaneous scalarization}. This effect was originally discussed by Damour and Esposito-Far\`ese \cite{Damour1993} in the context of a massless ($V(\phi) = 0$) scalar-tensor theory with conformal coupling $A(\phi) = \exp (\beta \phi^2/2)$. In this case it was noted that above a certain density threshold there is an abrupt increase in the strength of the NS scalar field content (or, more precisely, of its scalar charge), together with the appearance of new equilibrium solutions. The spontaneous scalarization effect has been generalized and analyzed from various perspectives; we point to Ref.~\cite{Berti2015} for a collection of relevant references. 


In the top panel of Fig.~\ref{fig:dil1} we show the central value of the scalar field as a function of the star's central density in the dilaton model defined by Eq.~(\ref{eq:dilaton}). As the central density increases and a pressure-dominated core begins to form and grow, we observe a sudden amplification of the scalar field content, together with sequential jumps in the number of equilibrium solutions. Figure \ref{fig:fieldprofileDil} shows the scalar field profiles for the eleven solutions found with a central density of $\tilde{\rho}_c = 10.0 \; \rho_\textrm{nuc}$. Contrary to the case of the chameleon model, in which the scalar field amplification in stars with pressure-dominated cores was mild (cf.~Fig.~\ref{fig:fieldprofile}), we see that here the scalar field rises orders of magnitude above the asymptotic value (which, for the theory parameters used in the plot, is $\phi_\infty \approx 3.64 \times 10^{-4} M_\textrm{Pl}$). This is typical of the scalarization phenomenon. Indeed, it is not surprising that the environmentally-dependent dilaton model may display this effect, since the coupling function (\ref{eq:dilaton}) resembles the model of Damour and Esposito-Far\'ese for small values of the scalar field.

From Fig.~\ref{fig:fieldprofileDil} we see that the new solutions typically exist in pairs, for which structural properties (such as mass and radius) are nearly identical and the scalar field profile is almost the same but with an opposite sign. This stems from the fact that the model (\ref{eq:dilaton}) becomes invariant under reflection, $\phi \to - \phi$, if $V_0 \to 0$, and the natural values for $V_0$ are quite small. The fact that these solutions have a different number of nodes is simply a consequence of the fact that the asymptotic value of the scalar field (such that $dV/d\phi|_{\phi_\infty} = 0$) is positive, and the solution with a negative central value of the scalar field has to cross zero one more time to reach it. 

In the bottom panel of Fig.~\ref{fig:dil1}, we show the total mass as a function of the star's central density. Only the first branches of scalarized solutions are clearly distinguishable, while the high-$n$ solutions accumulate around the GR equilibrium curve. For the branch of solutions with $n=0$, we find that the maximum mass decreases roughly $1.3 \%$ with respect to the GR value. This is a small decrease, comparable to what we found in the chameleon model (cf.~Fig.~\ref{fig:massradius}), although the scalar field activation seems more dramatic in this case.

\begin{figure}[th]
\includegraphics[width=9cm]{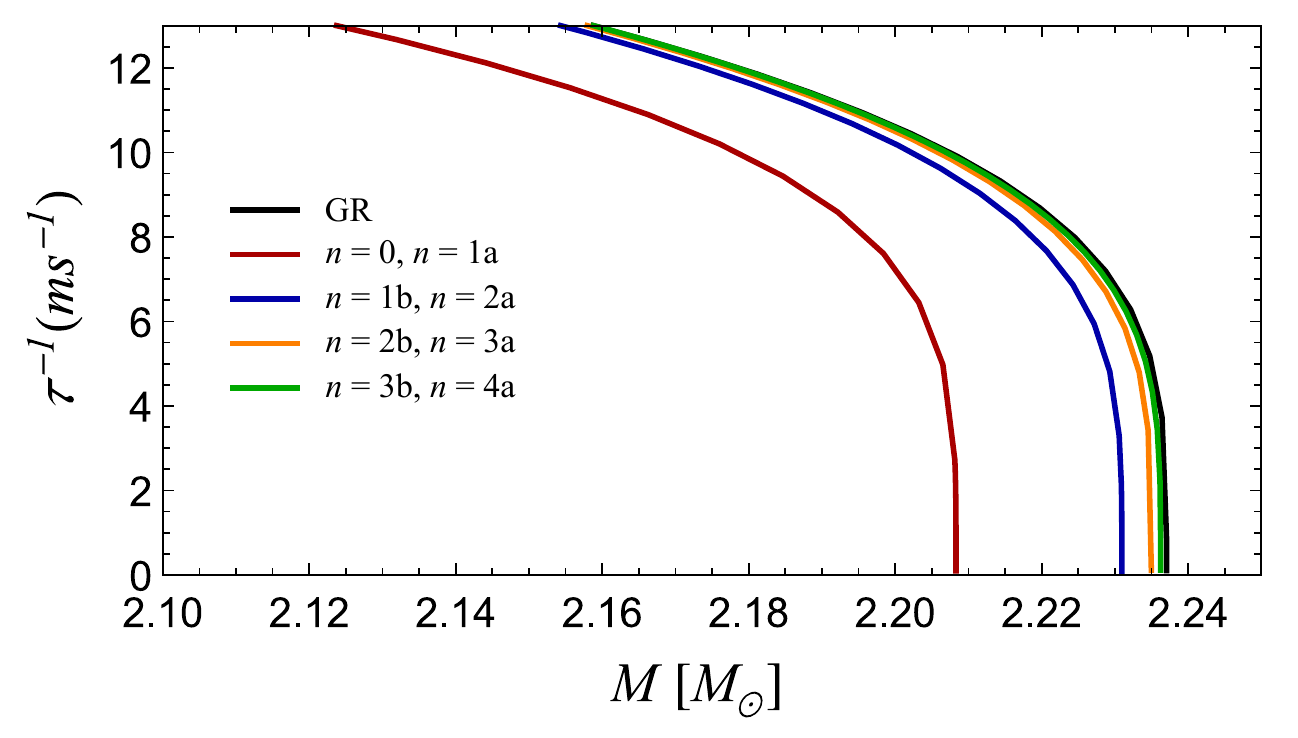}
\caption{Inverse of the instability timescale as a function of total mass for NS solutions in GR and in the dilaton model. The EoS and model parameters are the same as in Fig.~\ref{fig:dil1}. For readability, we display only the first families of solutions.
For a given number of nodes $n>0$, two branches of solutions are typically found: We denote by ``a'' (``b'') the branch with the largest (smallest) value of $|\phi_c|$. The instability timescale for solutions with $n = 0$ and $n = 1$a (and so on; see plot legend) is indistinguishable in the plot. For each of these branches, the maximum mass configuration presents a marginally stable mode, while denser configurations are radially unstable.}
\label{fig:modesdil}
\end{figure}

An important question that follows is whether and which equilibrium solutions found in this model are stable. First, we consider their dynamical stability under radial perturbations, searching for unstable radial modes (with time dependence $\exp(\Omega t)$, $\Omega >0$), as described in Sec.~\ref{sec:perturbations}. Figure.~\ref{fig:modesdil} shows the inverse of the instability timescale $\tau^{-1} = \Omega$ as a function of the total mass, for the first few branches of solutions displayed in Fig.~\ref{fig:dil1}. For each of these branches, we find that the maximum mass configuration is marginally stable (exhibiting a mode with $\Omega \approx 0$) and unstable modes exist for solutions with larger central densities. We find no evidence of unstable modes for solutions with central densities smaller than that of the maximum mass configuration. Branches where the scalar field has a large number $n$ of nodes have an instability timescale very close to the GR values; the same happened for their equilibrium properties (see Fig.~\ref{fig:dil1}).

In order to further access the stability of the new solutions present in the environmentally-dependent dilaton model, we have computed the binding energy $E_b \equiv M_b - M$ as a function of the baryon mass $M_b$ for different branches of equilibrium solutions. For a fixed value of $M_b$, the solution with $n=0$ and the one with $n=1$ and the largest value of $|\phi_c|$, have total masses that are indistinguishable within our numerical accuracy. We find that they are energetically favored over the remaining solutions, since they possess the highest binding energy for fixed baryon mass. 


As a final remark, it is worthwhile to mention that, in principle, the validity of the dilaton model given by Eqs.~(\ref{eq:dilaton}) and (\ref{eq:fieldtransf}) is restricted to values of $\phi$ close to zero. Although the field is indeed small inside equilibrium solutions for low density stars (cf.~Fig.~\ref{fig:dil1}), it can acquire large values inside scalarized solutions, which could call for a study of a more complete model.

\section{Final discussion} \label{sec:conclusion}

Scalar-tensor theories of gravity offer an interesting framework for cosmology, since the new scalar degree of freedom, when active at the largest scales in the universe, could help to drive its accelerated expansion. On the other hand, their physical viability relies on their ability to screen off the scalar-mediated fifth force in solar system scales. This is often accomplished by harnessing the coupling to the trace $T$ of the energy-momentum that is natural in these theories, and adjusting the model functions so that the scalar fifth force is suppressed in high density environments, but let free to operate in the low-density cosmological domain. 

On the other hand, some realistic EoS for nuclear matter predict that $T$ may change sign in the core of the most massive neutron stars, as pressure overtakes the rest-mass density as the main contributor for $T$. In this work we have explored some of the consequences of such a change for modified theories of gravity with screening mechanisms. In particular, we considered the chameleon and environmentally-dependent dilaton models, and studied equilibrium solutions describing neutron stars, as well as their dynamical stability under radial perturbations. 

In the chameleon model we find that the scalar field may become partially unscreened inside NSs with pressure dominated cores, leading to imprints in observable quantities such as their masses and radii. A stability analysis reveals that some of these solutions are stable under radial perturbations. Our results for the chameleon model are condensed in Figs.~\ref{fig:fieldprofile}, \ref{fig:massradius}, and \ref{fig:modescham}, and the accompanying discussion. 

For the case of an environmentally-dependent dilaton, we find that the existence of NSs with pressure-dominated cores leads to even more dramatic effects: An increase by some orders of magnitude in the scalar field content, together with the appearance of numerous branches of equilibrium solutions. The behavior is typical of the spontaneous scalarization phenomenon. A radial stability analysis reveals that some of the branches of solutions contain configurations that are stable under radial perturbations. Our results for the environmentally-dependent dilaton model are condensed in Figs.~\ref{fig:dil1}, \ref{fig:fieldprofileDil}, and \ref{fig:modesdil}, and the accompanying discussion.

Current constraints on modified theories of gravity displaying screening effects have mainly come from high-precision terrestrial experiments (see, e.g.,~Refs.~\cite{Upadhye2012a,Hamilton2015,Rider2016,Li2016}), and astrophysical observations in possibly unscreened environments, such as some dwarf galaxies \cite{Jain2011,Cabre2012,Jain2013,Vikram2018}. Possible tests coming from neutron star observations have considered pulsating sources \cite{Silvestri2011} or invoked the time-dependence of the scalar field background in the galaxy \cite{Brax2014}, but tend to be superseded by terrestrial and other astrophysical constraints. Our work suggests that, if the EoS for nuclear matter is such that stars with pressure-dominated cores are indeed found in Nature \footnote{See discussion in Refs.~\cite{Mendes2015,Podkowka2018} on how likely this is given the current understanding of the nuclear EoS.}, these could be promising sources of complementary constraints to modified theories of gravity with screening mechanisms. Our main goal was to demonstrate the existence of new phenomenology for such stars; we leave the investigation of actual constraints to the parameter space of relevant models for future studies.

Here we have focused on structural properties of NSs, such as their masses and radii. We showed, for instance, that the maximum NS mass can decrease with respect to GR by a few percent in the models we considered. An interesting development of the present work is to extend the study of radial perturbations---explored here with focus on stability issues---to the stable case, describing NS oscillations. In the Newtonian context, it was recently suggested \cite{Saltas2019} that helioseismology may be a good probe for fifth force effects. In the relativistic scenario, it was shown \cite{Mendes2018} that, in some scalar-tensor theories that exhibit the scalarization effect, NS oscillation frequencies can differ significantly from GR even when structural properties are similar, due to the presence of new families of modes (see also Ref.~\cite{blazquezsalcedo2020}). In particular, this could leave signatures in the post-merger waveform of a binary NS merger (see Ref.~\cite{Sagunski2018a} for an example), which could be observable by next-generation gravitational wave detectors \cite{TorresRivas2019}.

\acknowledgments
This work was partially supported by the National Council for Scientific and Technological Development -- CNPq.



\bibliography{library}

\end{document}